# IMPROVED Ti II log(gf) VALUES AND ABUNDANCE DETERMINATIONS IN THE PHOTOSPHERES OF THE SUN AND METAL-POOR STAR HD 84937

M. P. Wood[1], J. E. Lawler[1], C. Sneden[2], and J. J. Cowan[3]

[1]Department of Physics, University of Wisconsin, Madison, WI 53706; mpwood@wisc.edu, jelawler@wisc.edu

[2]Department of Astronomy and McDonald Observatory, University of Texas, Austin, TX 78712; chris@verdi.as.utexas.edu

[3]Homer L. Dodge Department of Physics and Astronomy, University of Oklahoma, Norman, OK 73019; cowan@nhn.ou.edu

# ABSTRACT

Atomic transition probability measurements for 364 lines of Ti II in the UV through near IR are reported. Branching fractions from data recorded using a Fourier transform spectrometer and a new echelle spectrometer are combined with published radiative lifetimes to determine these transition probabilities. The new results are in generally good agreement with previously reported FTS measurements. Use of the new echelle spectrometer, independent radiometric calibration methods, and independent data analysis routines enables a reduction of systematic errors and overall improvement in transition probability accuracy over previous measurements. The new Ti II data are applied to high resolution visible and UV spectra of the Sun and metal-poor star HD 84937 to derive new, more accurate Ti abundances. Lines covering a range of wavelength and excitation potential are used to search for non-LTE effects. The Ti abundances derived using Ti II for these two stars match those derived using Ti I and support the relative Ti/Fe abundance ratio versus metallicity seen in previous studies.

1. INTRODUCTION

Stellar abundances and their relative values as a function of metallicity ([Fe/H]) provide valuable information on the astrophysical origin of the chemical elements and the nucleosynthetic history of the Galaxy[1].  There exist fairly comprehensive databases (e.g., the NIST Atomic Spectra Database[2] and the Vienna Atomic Line Database[3]) of the atomic transition probabilities necessary for stellar abundance studies.  Studies of iron (Fe)-group abundances in metal-poor stars have found unexpected trends in relative abundance ratios verses metallicity (McWilliam et al. 1995a & 1995b, McWilliam 1997, Westin et al. 2000, Cowan et al. 2002, Sneden et al. 2003, Cayrel et al. 2004, Barklem et al. 2005, Lai et al. 2008, Bonifacio et al. 2009, Roederer 2009, Suda et al. 2011, Yong et al. 2013).  Some [X/Fe] trends cover ±1 dex from solar metallicity ([Fe/H] ≡ 0) down to metallicities of -4 (e.g., Figure 12 of McWilliam 1997).  These relative abundance trends are not compatible with current models of Fe-group yields from supernova in the early Galaxy.

While it was common practice to attribute such anomalies to inaccurate laboratory data, recent advances have essentially eliminated order of magnitude (1 dex) errors in experimental atomic transition probabilities.  The use of modern techniques, including laser induced fluorescence (LIF) for radiative lifetime measurements combined with Fourier transform spectrometer (FTS) measurements of emission branching fractions, has significantly improved the quality of experimental transition probabilities.  However, the possibility remains that inaccurate laboratory data still contribute to the observed Fe-group abundance trends.  In order to provide the most accurate abundances it is best to use lines that are weak enough in the photosphere of the star of interest to avoid saturation. In studies covering a wide range of metallicity (>2 dex), this necessitates using many lines covering a range of excitation potential (E.P.) and log(*gf*) values. In higher metallicity stars, one uses weaker lines with large E.P. in the first spectra (neutral atoms are a minor ionization stage in stars of interest). At lower metallicities, one must switch to stronger lines with lower E.P. and possibly to second spectra lines (singly ionized atoms are the dominant ionization stage), assuming suitable second spectra lines exist in the wavelength region being analyzed. The strength of these high and low E.P. lines can vary by orders of magnitude, making it difficult to accurately measure both with small uncertainties. This, combined with the possibility of using lines from more than one ionization stage, introduces the potential for transition probability uncertainties to affect the abundances. A second possible cause of the unexpected trends is that the 1D/LTE (one-dimensional/local thermodynamic equilibrium) photospheric models traditionally used to determine abundances are failing in metal-poor stars of interest (e.g. Asplund 2005). Metal-poor giants are favored in these studies to provide a large photon flux for high signal-to-noise (S/N), high resolution spectra. The low density atmospheres of giant stars combined with the reduced electron pressue from the low metal content results in lower collision rates, which might cause departures from LTE. The two possible explanations detailed above for the observed trends can be investigated by improving atomic transition probabilities in the Fe-group. If the trends are caused by 3D/non-LTE effects in giant metal-poor stars, one way forward is to map anomalous abundance results for various spectral lines covering a range of E.P. and wavelength in a wide range of stellar types. A third possible cause of the relative Fe-group abundance trends with metallicity is that models for supernova yields of Fe-group elements in the early Galaxy are inaccurate or incomplete. If the unexpected trends persist, even after improving transition probabilities and conducting targeted searches for non-LTE effects, it would be a strong indication that this third "nuclear physics" explanation is likely correct. In that case, models of Fe-group element production in the early Galaxy would need to be reexamined.

---

[1] We adopt standard spectroscopic notations.  For elements X and Y, the relative abundances are written [X/Y] = $\log_{10}(N_X/N_Y)_{star} - \log_{10}(N_X/N_Y)_\odot$.  For element X, the "absolute" abundance is written $\log \varepsilon(X) = \log_{10}(N_X/N_H)$.  Metallicity we be considered equivalent to the [Fe/H] value.
[2] Available at http://physics.nist.gov/PhysRefData/ASD/lines_form.html and http://physics.nist.gov/cgi-bin/ASBib1/Fvalbib/search_form.cgi.
[3] Available at http://www.astro.uu.se/~vald/php/vald.php.

An effort is underway to reduce transition probability uncertainties of selected neutral and singly-ionized Fe-group lines. The work of Den Hartog et al. (2011) focuses on sets of Mn I and Mn II lines useful for abundance determinations that span a range of E.P. and wavelength. The multiplets individually cover small wavelength ranges and/or are Russell-Saunders (LS) multiplets and it is therefore possible to reduce the log($gf$) uncertainties to 0.02 dex with $2\sigma$ confidence. While these small uncertainties are difficult to achieve and are practical only under favorable conditions, such small uncertainties are not required to detect non-LTE effects in metal-poor stars. Using these Mn I and Mn II data on selected metal-poor stars, effects of 0.5 to 1 dex are found confined to Mn I resonance lines connected to the ground level (Sobeck et al. 2013, in preparation).

As in our previous work on Ti I (Lawler et al. 2013; hereafter L13), this work on Ti II uses a broader approach than that of Den Hartog et al. (2011). Emission branching fraction measurements are attempted for every possible line connecting to the 42 odd-parity upper levels with lifetime measurements by Bizzarri et al. (1993). The result is a set of 364 log($gf$) values covering a wide range of E.P. and wavelength, with uncertainties ranging from 0.02 dex for dominant branches to ~0.10 dex for weak branches widely separated in wavelength from the dominant branch(es). The uncertainties on dominant branches are primarily from the lifetime measurements while the uncertainties on weak branches, which are typically the most important for abundance determinations, result primarily from the branching fraction measurements.

Branching fraction uncertainties are dominated by systematic effects which are often difficult to quantify, and this provides a major motivation for the construction of the new echelle spectrometer described in Section 3.

Using the new Ti II data, the Ti abundances in the photospheres of the Sun and metal-poor star HD 84937 are determined using many lines covering a range of E.P., wavelengths, and log(*gf*) values to search for non-LTE effects. A deviation toward lower abundance in HD 84937 is confined to wavelengths near the Balmer limit, consistent with the results of L13. This effect may be due in part to overpopulation of the n=2 levels of H. Lines of Ti II outside the Balmer continuum wavelength region yield consistent abundance values with no detectable dependence on wavelength, E.P., or line strength. The Saha ionization balance of Ti in HD 84937 is confirmed by good agreement between log($\varepsilon$(Ti II)) from this work with log($\varepsilon$(Ti I)) from L13. The [Ti/Fe] trend with metallicity observed in earlier studies of metal-poor stars is also supported in this study.

## 2. FOURIER TRANSFORM SPECTROMETER DATA

As in much of our previous branching fraction work, we use data from the 1.0 meter FTS previously at the National Solar Observatory (NSO) on Kitt Peak for this study of Ti II branching fractions. The NSO 1.0 meter FTS has a large etendue, a limit of resolution as small as 0.01 cm$^{-1}$, wavenumber accuracy to 1 part in 10$^8$, broad spectral coverage from the UV to IR, and the ability to record a million point spectrum in 10 minutes (Brault 1976). However, the NSO 1.0 meter FTS is no longer in operation, and this provides further motivation for the construction of a new echelle spectrometer as described in the next section.

In order to obtain high quality branching fractions, one must determine calibrated line intensities from multiple spectra. Overlapping visible-UV and IR spectra of lamps at high currents provide adequate S/N on very weak branches to all known lower levels. In addition, it is essential to have visible-UV spectra of the lamps operating at low currents in which the dominant branches are optically thin. Table 1 lists the 14 FTS spectra used in this Ti II branching fraction analysis. All 14 spectra, raw interferograms, and header files are available in the NSO electronic archives[4]. The majority of these spectra are from lamps operating at high currents, and some of these spectra have optical depth problems on the dominant branches. As described in the next section, a new echelle spectrometer is used to record additional Ti hollow cathode discharge (HCD) spectra to address the optical depth issues.

The determination of an accurate relative radiometric calibration is critical to a branching fraction experiment. As described in L13, the Ar I and Ar II line calibration method is used to calibrate the spectra in Table 1. This technique, which is internal to the HCD Ti/Ar lamp spectra, is our preferred method for FTS calibration. It is based on a comparison of well-known branching ratios for sets of Ar I and Ar II lines to the intensities measured for the same lines. Sets of Ar I and Ar II lines have been established for this purpose in the range from 4300 to 35000 cm$^{-1}$ by Adams & Whaling (1981), Danzmann & Kock (1982), Hashiguchi & Hasikuni (1985), and Whaling et al. (1993).

## 3. ECHELLE SPECTROMETER DATA

A powerful new echelle spectrometer for laboratory astrophysics at the University of Wisconsin is used in this work on Ti II. In addition to the optical depth problems mentioned above, the desire to map non-LTE effects on Fe-group abundances in metal-poor stars provides further motivation for the development of this new echelle spectrometer. As described in L13, in the photospheres of metal-poor stars of interest, >90% of Fe-group atoms are singly ionized, and of those ions ≥75% are in the ground or low metastable levels. Since the ground and low metastable ion levels are the primary population reservoir, lines connecting to these levels cannot be significantly out of equilibrium and as a result provide the best abundance determinations to map non-LTE effects in metal-poor stars. Strong branches connecting to these levels of the singly-ionized species can still be affected by resonance scattering, and therefore the weak lines connecting to these levels represent the "gold standard" lines for obtaining the most accurate abundances. Accurate transitions probabilities for these second spectra lines, most of which are in the UV, are thus very important. FTS instruments suffer from multiplex noise in which the quantum statistical (Poisson) noise from all spectra features, including strong visible and near-IR branches, is smoothly redistributed throughout the entire spectrum. Often, as the lamp current is reduced, weak lines become comparable to the multiplex noise before the dominant branches from the common upper level become optically thin. A dispersive spectrometer, which is free from multiplex noise, can provide adequate S/N on these astrophysically important weak UV lines.

---
[4] Available at http://nsokp.nso.edu/

The design and performance of the new echelle spectrometer, including a detailed aberration analysis, is described in detail by Wood & Lawler (2012). The spectrometer has a resolving power of 250,000 during normal data acquisition with broad spectral coverage, superb UV sensitivity, and great flexibility in wavelength from the UV to the silicon detector limit in the IR. Compared to a FTS, the most significant disadvantage is the reduced accuracy and precision of the wavenumber calibration while the most significant advantage is the absence of multiplex noise. The echelle spectrometer is radiometrically calibrated using continuum standard lamps.

These lamps allow the echelle spectrometer to provide good branching fraction measurements on Ti II lines down to 2300 Å since the lamps are calibrated to shorter wavelengths compared to the Ar I and Ar II calibration of the FTS described in Section 2.

In addition to the 14 FTS spectra in Table 1, the 48 CCD frames of Ti HCD spectra listed in Table 2, recorded with the echelle spectrometer, are part of this Ti II branching fraction study. All of the spectra listed in Tables 1 and 2 are also used by L13 for branching fraction measurements on Ti I. That work represents the first comprehensive branching fraction analysis for Ti I. In contrast, there are already two sets of published Ti II branching fraction measurements from FTS data (Bizzarri et al. 1993, Pickering et al. 2001). Even with the previously published work on Ti II, improvement of the log(*gf*) values is achieved. Systematic uncertainties dominate most branching fraction measurements and are difficult to control and reliably estimate. This work provides an assessment of the reproducibility of the branching fraction measurements using different spectrometers and radiometric calibration methods, and thus provides a better estimate of the systematic uncertainties. This work also expands upon the earlier results by including measurements of additional lines.

## 4. Ti II BRANCHING FRACTIONS

All possible transition wavenumbers between known energy levels of Ti II from Saloman (2012) that satisfy both the parity change and $|\Delta J| \leq 1$ selection rules are computed and used during analysis of the FTS and echelle data. (Transitions which violate these two selection rules are suppressed by a factor of $\sim 10^6$ and are not typically useful for stellar abundance studies). While many important Fe-group transitions violate the $|\Delta L| \leq 1$ and $\Delta S = 0$ selection rules of LS or Russell-Saunders coupling, the parity change and $|\Delta J|$ selection rules are always obeyed for electric dipole transitions. Since we are able to make measurements on branching fractions as weak as 0.0001, systematic errors from missing branches are negligible. Ti has five naturally occurring isotopes, two of which have hyperfine structure due to a non-zero nuclear spin. However, Ti II lines are rather narrow in the FTS data and component structure is neglected for this study.

Branching fraction measurements are attempted for all 42 odd-parity upper levels with lifetime measurements reported by Bizzarri et al. (1993) and are completed for 39 of those levels. The levels for which branching fractions could not be completed have a strong branch with a severe blending problem. As in L13, thousands of possible spectral line observations in both FTS and echelle spectra are analyzed during the study of lines from the 39 upper levels. Integration limits and nonzero baselines are set "interactively" during data analysis. Nonzero baselines are routinely needed during analysis of the echelle spectra as these are not dark signal corrected, and are occasionally used for FTS spectra when a line falls on the wing of a nearby dominant feature. To determine un-calibrated Ti II lines intensities, a simple numerical integration technique is used because of unresolved or weakly resolved isotopic and/or hyperfine structure. This same numerical integration routine is also used on selected Ar I and Ar II lines to establish a relative radiometric calibration of the FTS spectra.

The procedure for determining branching fraction uncertainties is described in detail by Wickliffe et al. (2000) and is the same employed by L13. Branching fractions for weaker lines near the dominant line(s) tend to have uncertainties limited by the S/N. For widely separated lines from a common upper level, systematic uncertainties in the radiometric calibration are typically the most serious source of uncertainty. We use a formula for estimating this systematic uncertainty that is presented and tested extensively by Wickliffe et al. (2000). Redundant measurements from spectra with different discharge conditions allow the identification of blended lines. The final branching fraction uncertainties are primarily systematic and it is thus difficult to say whether these uncertainties represent 1σ or 2σ error bars.

Measured branching fractions from the combined FTS and echelle spectra are compared with the results of Bizzarri et al. (1993) and Pickering et al. (2001) in Table 3. Only a "stub" version of Table 3 is included in the printed version of this article, with the full electronic or machine readable version available online. This table provides and assessment of the reproducibility in the branching fraction measurements. The use of different spectrometers and radiometric calibrations allows for an assessment and reduction of systematic uncertainty.

## 5. Ti II TRANSTION PROBABILITIES AND COMPARISON TO EARLIER MEASUREMENTS

Branching fractions reported in Table 3 are normalized using radiative lifetime measurements (Bizzarri et al. 1993) to determine absolute transition probabilities for the 364 Ti II lines in Table 4. Air wavelengths are computed using the standard index of air (Peck and Reeder 1972). The full electronic or machine readable version of Table 4 is available online, while a stub version is included here to guide the reader. Some problem lines that are too weak to have reliable S/N, have uncertain classification, or are too seriously blended have been omitted Table 4. This can be seen by summing all transition probabilities from a given upper level and comparing the sum to the inverse of the upper level lifetime, and for most levels the sum is between 90% and 100% of the inverse lifetime. Branching fraction uncertainties are combined in quadrature with lifetime uncertainties to determine the transition probability uncertainties given in Table 4.

Figure 1 is a comparison of 84 lines in common between this work and the work of Bizzarri et al. (1993). Error bars represent the log(*gf*) uncertainty from this work and the Bizzarri work combined in quadrature. We note that ~83% of the log(*gf*) values in common agree within their combined uncertainties. Figure 2 shows the same 84 lines in common plotted as a function of the log(*gf*) measured in this work to emphasize that the most significant scatter is confined to the weaker lines. Figure 3 is a comparison of 299 lines in common between this work and the results of Pickering et al. (2001). Again, the error bars represent the log(*gf*) uncertainty of the two works combined in quadrature. There is generally good agreement, with approximately 70% of the log(*gf*) values in common agreeing within combined uncertainties. Figure 4 show the same 299 lines in common plotted as a function of the log(*gf*) measured in this study.

While some significant outliers exist in Figures 1-4, those lines are well positioned and well resolved in our data and we are confident in our branching fraction results. This work has the advantage of improved Ti II energy levels from Saloman (2012) which help with many ambiguous line classifications and possible blends. Users can convert our results to log(*gf*) values based on a different source of branching fractions or based on an average of branching fraction measurements (e.g. in Table 3) by adding log(BF$_{new}$/BF$_{Wood}$) to our reported log(*gf*) values.

## 6. THE TITANIUM ABUNDANCE IN THE SOLAR PHOTOSPHERE

We use our new Ti II transition probability data to re-determine the solar photospheric Ti abundance. We follow without change the procedures used by L13 in their application of new Ti I transition probabilities to the solar spectrum.

We begin by computing approximate relative strengths of all 364 Ti II lines of this study. As in previous papers by our group, we compute relative absorption strengths of Ti II lines as

$$\mathrm{STR} \equiv \log(gf) - \theta\chi$$

with the $\log(gf)$ given in Table 4, excitation energies in units of eV, and inverse temperature $\theta = 5040/T$. The relatives strengths computed in this simple manner strictly apply only to a single species, and highlight the well-known (e.g. Gray 2008) strong dependence of the absorption line strength on two factors: the transition probability and the Boltzmann excitation factor. For simplicity we adopt here $\theta = 1.0$, which is close enough to the temperature of solar photospheric line forming layers for these illustrative calculations. The STR values as a function of wavelength are plotted in Figure 5. As in other plots of this type shown in previous papers, one can easily see that the strongest lines are at shorter wavelengths, and mostly lie below 4000 Å. In Figure 5 we have drawn a horizontal line at STR = –5, and estimate that a line of this strength will produce an absorption feature in the solar spectrum with reduced width $\log(RW) = \log(EW/\lambda) \sim -6$ (e.g., lines with $EW \sim 0.5$ mÅ at $\lambda = 5000$ Å). Given the excellent spectral resolution and S/N of available solar spectra, $\log(RW) \sim -6$ is near the limit of weak-line detectability. Only about a dozen lines in our study have STR $<\sim -5$, and essentially all Ti II lines should have significant absorption in the solar photospheric spectrum. Indeed, more than half of these lines should be "strong", defined here as having STR $> -3$, or more than 100 times stronger than the detection limit.

We therefore assess the potential utility of all 364 Ti II lines for deriving a new solar Ti abundance. We employ the Delbouille et al. (1973) center-of-disk spectrum as well as the Moore et al. (1966) solar line identification compendium to eliminate Ti II transitions whose photospheric absorptions are undetectably weak (only a few, given the discussion above) or too seriously blended with various atomic and/or molecular species. This process leaves about 60 Ti II lines which are potentially useful for solar abundance analysis.

We generate synthetic spectra for these transitions to compare with the Delbouille et al. (1973) solar spectrum. We produce multiple syntheses with different assumed Ti abundances, and smooth them with Gaussian broadening functions to match the combined effects of the very narrow spectrograph instrumental profile (resolving power, $R = \lambda/\Delta\lambda > 5\times10^5$) and solar macroturbulence. Development of the line lists is described by L13 in detail. Briefly, this procedure begins with the line compendium of Kurucz (2011)[5] for a small spectral region surrounding a Ti II line of interest, adopts recent lab transition probabilities of lines in this wavelength interval (including Ti I results from L13 and Ti II from this study), and adjusts the *gf* values of lines with no lab data to best match the solar spectrum. To compute synthetic solar spectra we use the Holweger & Müller (1974) empirical model photosphere, these line lists, and the current version of the LTE line analysis code MOOG[6] (Sneden 1973). These detailed synthetic/observed spectrum comparisons force us to eliminate more potential Ti II lines because they prove to be unacceptably contaminated by other absorption features.

In Figure 6 we show examples of the observed/synthetic spectral matches. The assumed Ti abundance for each synthesis is explained in the figure caption. The three Ti II lines chosen for display are mostly unblended. The order of presentation from panels (a)–(c) is in decreasing wavelength but also increasing line strength. In particular, the 3263.7 Å line in panel (c) is very strong (STR ~ -2.3) and therefore large changes in Ti abundance do not produce large absorption differences in this line. Abundances from such strong lines should be treated with caution.

---

[5] http://kurucz.harvard.edu/linelists.html

[6] Available at http://www.as.utexas.edu/~chris/moog.html

From a final set of 43 usable Ti II lines we re-determine the solar Ti abundance. The parameters of these lines are listed in Table 5. The mean abundance is $<\log \varepsilon(Ti)> = 4.979 \pm 0.005$ with $\sigma = 0.032$. These abundances are plotted as a function of wavelength in Figure 7. The mean abundance from Ti II is in excellent agreement with the abundance derived from Ti I by L13: $<\log \varepsilon(Ti)> = 4.973 \pm 0.003$ with $\sigma = 0.040$. As seen in Figure 7, the atomic structures of Ti I and Ti II inevitably lead to more detectable neutral-species transitions at longer wavelengths than those of the ion (similar to other Fe-group elements). However, there is significant overlap of useful Ti I and Ti II lines in the 4400-5400 Å region, and for these lines the mean abundances from neutral and ion transitions agree well. To echo the conclusion of L13, our new photospheric Ti abundance is also in reasonable accord with the recommended solar photospheric abundances of $4.95 \pm 0.05$ (Asplund et al. 2009) and $4.90 \pm 0.06$ (Lodders, Palme, & Gail 2009)

## 7. THE TITANIUM ABUNDANCE OF HD 84937

HD 84937 is a bright metal-poor main-sequence turnoff star that has enjoyed many abundance analyses over the past couple of decades. As detailed in L13, the atmospheric parameters effective temperature, surface gravity, metallicity, and microturbulent velocity appear to be well-determined. We adopt the values for these quantities recommended in that paper: $T_{eff}$ = 6300 K, log g = 4.0, [Fe/H] = –2.15, and $v_t$ = 1.5 km s$^{-1}$. We search for Ti II lines appropriate for abundance analysis in the optical wavelength *ESO VLT UVES* and vacuum-UV *HST/STIS* spectra (obtained under proposal #7402 by R. C. Peterson) as described in L13. The lower metallicity of HD 84937 yields more Ti II lines useful for abundance determinations, both because very saturated lines in the solar photosphere become much weaker (and more sensitive to abundance changes) and contamination from other species is significantly reduced. We show this in Figure 8 by displaying the same lines in HD 84937 that are in the solar photospheric spectra of Figure 6. These lines are uncomplicated and can be synthesized without difficulty.

In Table 6 we list line parameters and derived abundances for all 147 Ti II transitions used for HD 84937. The mean abundance is $\langle\log \varepsilon(Ti)\rangle = 3.081 \pm 0.007$ with $\sigma = 0.087$. This value is a slightly lower than L13's result for Ti I: $\langle\log \varepsilon(Ti)\rangle = 3.122 \pm 0.007$ ($\sigma = 0.054$; 54 lines). More importantly, as shown in Figure 9, there is a trend with wavelength for abundances derived using Ti II lines. For 44 Ti II lines with wavelengths $\lambda > 3800$ Å we obtain $\langle\log \varepsilon(Ti)\rangle = 3.123 \pm 0.006$ ($\sigma = 0.037$), in excellent agreement with the abundance from Ti I. The single-line standard deviations $\sigma$ of the two species' abundances are similarly small and likely represent the true "measurement errors". In contrast, for 103 lines with wavelengths $\lambda < 3800$ Å the mean abundance is lower and the standard deviation higher: $\langle\log \varepsilon(Ti)\rangle = 3.063 \pm 0.009$ ($\sigma = 0.096$). Note also that from 16 lines in the vacuum-UV ($\lambda < 3000$ Å) we obtain abundances in agreement with the Ti I and long wavelength Ti II results: $\langle\log \varepsilon(Ti)\rangle = 3.14 \pm 0.02$ ($\sigma = 0.07$). Thus the large line-to-line scatter and low mean abundance appear to arise mostly in the wavelength interval 3000–3800 Å. To proceed further, we first briefly consider four possible but unlikely resolutions to this problem.

(1) Incorrect effective temperature: as discussed by L13, $T_{eff}$ is very well-determined from both photometry and spectroscopy. Additionally, nearly all the Ti II lines of this study have a small range of E.P. (0–2.5 eV), muting the Boltzmann factor response to small temperature changes.

(2) Incorrect gravity: HD 84937 has a well-determined distance, which leads to a robust gravity estimate (see L13). Here, we emphasize that for lines with λ > 3800 Å, abundances derived from Ti II and Ti I (L13) are in excellent accord.

(3) Incorrect microturbulence: this parameter has not been independently assessed in this study. However, most Ti II lines are not extremely strong in HD 84937. Measurements of lines used in our study from the 3400–3600 Å region yield <log(RW)> ≈ –5.0. Such lines are not strong enough to be very sensitive to adopted values of $v_t$. Nevertheless, we resynthesize the nearly thirty Ti II lines with the lowest abundances using a microturbulence decreased from $v_t$ = 1.5 km s$^{-1}$ to 1.0 km s$^{-1}$. Little change is found in derived Ti abundances, with the increases in log ε ranging from 0.01 to 0.15, and the strongest lines exhibiting the largest changes. The mean increase is ≈0.10, which is insufficient to completely erase the mean abundance difference between lines in this spectral region and those with λ > 3800 Å, and it does not significantly impact the line-to-line scatter. Over-interpretation of these simple calculations should be avoided, acknowledging that microturbulence here is a single-valued parameter attempting to describe a complex physical environment. For example, we have not attempted to introduce depth-dependence into this parameter (as explored in, e.g., Takeda et al. 2006, and references therein). In addition, we have not explored the application of 3D stellar atmospheres (e.g. Trampedach et al. 2013) to this issue, and we have ignored possible non-LTE effects in Ti line formation (Bergemann 2011). All of these issues deserve careful exploration in the future.

(4) Incorrect branching fractions: In Figure 10 we repeat the abundance versus wavelength plot of Figure 9, but this time divide the Ti II lines into those with small branching fractions (BF < 0.2) and those with larger ones. It is clear that the transitions with strong branching fractions dominate the low abundance points in this plot. However, we cannot blame the branching fractions here. Transition probability uncertainties are typically dominated by branching fraction uncertainties. Both the earlier Ti II study by Pickering et al. (2001) and the new lab study reported herein use the same set of radiative lifetimes (Bizzarri et al. 1993) to convert branching fractions into absolute transition probabilities. These lifetimes have uncertainties of ±5% (±0.02 dex) and the uncertainties are strongly supported by independent laser induced fluorescence measurements including a laser-fast beam measurement that is thought to be accurate to ±1.5%. A comparison of the branching fractions for lines plotted in Figure 10 with crosses (corresponding to BF > 0.2) from this work and from Pickering et al. reveals excellent agreement, with $<\log(BF_{Pickering}/BF_{Wood})> = 0.007 \pm 0.004$ ($\sigma = 0.022$) for the 25 of 26 lines in common to both studies. Stronger branching fractions are easier to measure because branching fractions are defined to sum to unity and thus uncertainty migrates to weaker branches.

In Figure 10 there is a slight (~ 0.1 dex) dip in abundance for both Ti II lines with BF < 0.2 and Ti I lines. This abundance dip in the near-UV was previously noted by L13. They derive a new Fe abundance for HD 84937 from a large number of Fe I lines, adopting transition probabilities from the NIST "critical compilation" of Fuhr & Wiese (2006)[7]. We extend the work of L13 to Fe II, and the results are shown in Figure 11. As in L13, we choose to include only Fe II lines with quality grade of "C" or better[8]. From all Fe II transitions we derive $<\log \varepsilon(Fe)> = 5.19 \pm 0.01$ ($\sigma = 0.06$; 113 lines), and we note that the UV transitions yield similar abundances to the long-wavelength ones. The abundance anomaly in the 3300–3700 Å is clearly seen in Fe I. Unfortunately, there are almost no detectable Fe II lines with high-quality transition probabilities in this wavelength region.

---

[7] Most of these data are available at http://www.nist.gov/pml/data/asd.cfm
[8] From Fuhr and Wiese (2006), the estimates of minimum accuracies for each grade of interest here are: A, ≤3%; B, ≤10%; C, ≤25%; D, ≤50%; and E, >50%.

Additional insight into the near-UV abundance dip can be gained by grouping the Ti II abundances by the lower level energy (E.P.) of the line. The ground and lowest energy levels are in the a$^4$F and b$^4$F terms, which include eight levels near or below 0.13eV, and the a$^2$F term comprising 2 levels near 0.59eV. These ten levels are stable (ground) or metastable levels which do not decay via electric dipole transitions. In a simple Boltzmann calculation, these 10 ground and low metastable levels account for 90% of the population of the whole Ti II species. In Figure 12 we again show an abundance versus wavelength plot, this time dividing the lines into those arising from levels of the three lowest terms and those arising from levels with higher E.P. Most Ti II transitions arising from the ten lowest levels occur in the 3000–3800 Å Balmer continuum region, yet their mean abundance is <log ε(Ti)> = 3.135 ± 0.007 (σ = 0.046; 47 lines). Clearly this is in excellent agreement with the means and standard deviations of Ti II lines at longer wavelengths and of Ti I lines. Moreover, the lines that we have successfully analyzed in the 3000–3800 Å region include examples that are very weak transitions (very sensitive to the Ti abundance) and very strong (saturated, less sensitive to abundance). Indeed, the transition probabilities for lines arising from the low-excitation group range over two orders of magnitude. Therefore attribution of the problem to line strength issues is unlikely to be correct.

The dip in Fe I abundance values in the Balmer continuum region is also seen and studied in detail in other metal-poor stars by Roederer et al. (2012). There is essentially no uncertainty in the photoionzation (bound-free) cross section for the n = 2 levels of H and the *f*-values of the Balmer (bound-bound) series are exact analytic values. The only possible uncertainty in the atomic data for H in the near-UV is due to the selection of a model for terminating the Balmer series at the ionization limit. Roederer et al. investigated several models for the n = 2 series termination, but the dip in Fe I abundance values in the near-UV persisted. The Ti II results in the near-UV are roughly consistent with the anomalous Fe I abundance values for HD 84937 in the near-UV determined by L13 and with the Fe I abundance values found by Roederer et al. in the near-UV spectra of four other metal poor stars. These results collectively suggest some extra continuum opacity in the near-UV, perhaps at least partially resulting from a non-LTE population of the n = 2 levels of H in these metal-poor stars. However, the fact that the Ti II abundance dip in the Balmer region has a dependence upon E.P., an effect not seen in Fe I by Roederer et al. (2012), suggests that other non-LTE effect are likely to factor in. Certainly studies of such issues are recommended for the future before definitive conclusions can be made.

However, the existence of an abundance dip in the near-UV should not obscure our overall conclusion: application of a standard model-atmosphere synthetic-spectrum analysis to many lines of Ti II, using the new accurate transition probabilities reported herein, and Ti I yields a consistent Ti abundance in HD 84937. We do not find significant departures from LTE in the ionization equilibrium, in good agreement with the assertion of Bergemann (2011). It will be of interest to now apply these transition data to the spectra of metal-poor giant stars.

## 8. SUMMARY

New accurate atomic transition probabilities for 364 lines of Ti II are reported. Branching fractions measured from archived FTS spectra and new spectra recorded with an echelle spectrometer are combined with LIF lifetime measurements to determine the transition probabilities. Overall good agreement is found in comparisons of these new data to previously reported Ti II transition probabilities. Using the new Ti II data, Ti abundances are determined in the photospheres of the Sun and the metal-poor star HD 84937 using lines covering a range of wavelength, E.P., and log($gf$) values. The abundance results for the Sun are in excellent agreement with those found by L13 using new Ti I data. In HD 84937, abundance values derived from Ti II lines with $\lambda > 3800$ Å have small scatter and are in good agreement with the results of L13. More line-to-line scatter and a lower mean abundance are found for shorter-wavelength lines. As in L13, this Ti II study yields evidence for the presence of extra photospheric continuum opacity near the Balmer limit. However, the variation of [Ti/Fe] with metallicity observed in previous studies of metal-poor stars is supported in this study.

ACKNOWLEDGEMENTS

The authors acknowledge the help of A. J. Fittante in early data analysis on this project. This work is supported in part by NASA Grant NNX10AN93G (J.E.L.) and NSF Grants AST-0908978 and AST-1211585 (C.S.). Some data for this project were obtained from the ESO Science Archive Facility under request numbers 073.D-0024 (PI: C. Akerman) and 266.D-5655 (Service Mode). Other data were from observations made with the NASA/ESA *Hubble Space Telescope*, obtained from the data archive at the Space Telescope Science Institute. STScI is operated by the Associated of Universities for Research in Astronomy, Inc. under NASA contract NAS 5-26555. The solar spectra were acquired from the BASS2000 solar data archive.

FIGURE CAPTIONS

Figure 1. Comparison of our (Wood) log(*gf*) values for 84 lines in common with Bizzarri et al. (1993). The central dotted line indicates perfect agreement. Error bars represent the Wood and Bizzarri uncertainties combined in quadrature.

Figure 2. The same 84 lines compared in Fig. 1 plotted as a function of our measured log(*gf*) value. The central dotted line and error bars have the same meaning as those in Fig. 1.

Figure 3. Comparison of our (Wood) log(*gf*) values for 299 lines in common with Pickering et al. (2001). The central dotted line indicates perfect agreement. Error bars represent the Wood and Pickering uncertainties combined in quadrature.

Figure 4. The same 299 lines compared in Fig. 3 plotted as a function of our measured log(*gf*) value. The central dotted line and error bars have the same meaning as those in Fig. 3.

Figure 5. Relative strengths of the Ti II lines considered in this study, plotted as a function of their wavelengths; see text for definitions of terms. The vertical blue line denotes the Earth's atmospheric cutoff wavelength; lines shortward of this wavelength are inaccessible to ground-based telescopes. The horizontal blue line denotes the approximate strengths of barely detectable Ti II lines: ones that have reduced widths log(*RW*) = –6. Red circles highlight those lines that are used in the solar abundance analysis.

Figure 6. Observed and synthetic spectra of three representative Ti II lines in the solar spectrum. The open circles represent every 4$^{th}$ point of the very high-resolution Delbouille et al. (1973) atlas. The lines in each panel represent five synthetic spectra: black color is for the Ti abundance for this line reported in Table 3; orange is for an abundance 0.3 dex larger; green is for an abundance 0.3 dex smaller; blue is for an abundance 0.6 dex smaller; and red is for a synthesis computed without any contribution from Ti.

Figure 7: Solar Ti abundances from Ti II lines (red; this study) and Ti I lines (blue; L13) are plotted against wavelength. The statistics are written in the figure legend. The horizontal red line is placed at the mean abundance derived from Ti II only, and the dotted lines are placed at ±1σ from this mean.

Figure 8. Observed and synthetic spectra of three Ti II lines in the HD 84937 spectrum. The transitions displayed here are the same ones that we selected for Figure 6. The open circles are from the *VLT UVES* observed spectrum, and the lines have the same meanings as those of Figure 6.

Figure 9: Ti abundances from Ti II lines (red open circles and green ×'s; this study) and Ti I lines (blue dots; L13) for HD 84937 are plotted against wavelength. The statistics are written in the figure legend. The horizontal red line is placed at the mean abundance derived from Ti II lines with λ > 3800 Å, and the dotted lines are placed at ±1σ from this mean. See text for a discussion of the split in Ti abundances into two groups divided at 3800 Å.

Figure 10: As in Figure 9, but here green × symbols designate strong branching-fraction Ti II transitions and red open circles designate weaker ones.

Figure 11: Fe abundances from Fe I and Fe II lines for HD 84937. The horizontal black line represents the mean of all Fe II line abundances, and the dotted lines represent ±1σ from this mean.

Figure 12: As Figure 9, but here the green × symbols designate transitions from the ground and nine lowest energy metastable levels, and red open circles designated transitions from higher excitation levels.

Table 1. Fourier transform spectra of a custom water-cooled Ti hollow cathode discharge (HCD) lamp. All were recorded using the 1 m FTS on the McMath telescope at the National Solar Observatory, Kitt Peak, AZ.

| Index | Date | Serial Number | Lamp Type | Buffer Gas | Lamp Current (mA) | Wavenumber Range ($cm^{-1}$) | Limit of Resolution ($cm^{-1}$) | Coadds | Beam Splitter | Filter | Detector[1] |
|---|---|---|---|---|---|---|---|---|---|---|---|
| 1 | 1991 Feb. 21 | 4 | Custom HCD | Ar-Ne | 240 | 7972 - 41160 | 0.042 | 8 | UV | | Mid Range Si Diode |
| 2 | 1989 Mar. 9 | 16 | Custom HCD | Ar | 770 | 7404 – 33520 | 0.042 | 3 | UV | WG295 | Super Blue Si Diode |
| 3 | 1989 Mar. 14 | 18 | Custom HCD | Ar | 900 | 7673 – 34328 | 0.042 | 6 | UV | WG295 | Super Blue Si Diode |
| 4 | 1991 May 1 | 4 | Custom HCD | Ar | 434 | 8460 - 43317 | 0.050 | 6 | UV | $CuSO_4$ | Mid Range Si Diode |
| 5 | 1991 May 3 | 16 | Custom HCD | Ar | 150 | 8460 - 43317 | 0.050 | 6 | UV | $CuSO_4$ | Mid Range Si Diode |
| 6 | 1991 May 3 | 17 | Custom HCD | Ar | 610 | 8460 - 43317 | 0.050 | 6 | UV | $CuSO_4$ | Mid Range Si Diode |
| 7 | 1989 Feb. 28 | 25 | Custom HCD | Ar | 820 | 8368 - 36310 | 0.042 | 6 | UV | WG295 | Super Blue Si Diode |
| 8 | 1992 July 28 | 1 | Custom HCD | Ar | 410 | 8334 - 45394 | 0.046 | 8 | UV | WG230 | Mid Range Si Diode |
| 9 | 1992 July 29 | 2 | Custom HCD | Ar | 140 | 8421 - 45507 | 0.046 | 2 | UV | WG230 | Mid Range Si Diode |
| 10 | 1992 July 29 | 3 | Custom HCD | Ar | 145 | 8421 - 45507 | 0.046 | 4 | UV | WG230 | Mid Range Si Diode |

| 11 | 1992 July 29 | 4 | Custom HCD | Ar | 148 | 8421 - 45865 | 0.046 | 8 | UV | WG230 | Mid Range Si Diode |
| 12 | 1991 May 1 | 1 | Custom HCD | Ne | 53 | 8460 - 43317 | 0.050 | 6 | UV | $CuSO_4$ | Mid Range Si Diode |
| 13 | 1991 May 1 | 2 | Custom HCD | Ne | 102 | 8460 - 43317 | 0.050 | 6 | UV | $CuSO_4$ | Mid Range Si Diode |
| 14 | 1991 May 1 | 3 | Custom HCD | Ne | 202 | 8460 - 43317 | 0.050 | 6 | UV | $CuSO_4$ | Mid Range Si Diode |

[1]Detector types include the Super Blue silicon (Si) photodiode and a Mid Range Si photodiode.

Table 2. Echelle spectra of commercial Ti HCD lamps and a custom water-cooled Ti HCD lamp.

| Index | Date | Serial Numbers[1] | Lamp Type[2] | Buffer Gas | Lamp Current (mA) | Wavelength Range (Å) | Resolving Power | Coadds | Expos. Time (s) |
|---|---|---|---|---|---|---|---|---|---|
| 21-23 | 2011 July 27 | 4, 8, 12 | Commercial HCD | Ne | 10 | 2200-3900 | 250,000 | 10 | 360 |
| 24-26 | 2011 July 28 | 4, 8, 12 | Commercial HCD | Ne | 15 | 2200-3900 | 250,000 | 10 | 360 |
| 27-29 | 2011 Aug. 12 | 4, 8, 12 | Commercial HCD | Ne | 7 | 2200-3900 | 250,000 | 20 | 180 |
| 30-32 | 2011 Aug. 9 | 4, 8, 12 | Commercial HCD | Ar | 10 | 2200-3900 | 250,000 | 10 | 360 |
| 33-35 | 2011 Sept. 1 | 4, 8, 12 | Custom HCD | Ar | 25 | 2200-3900 | 250,000 | 20 | 120 |
| 36-38 | 2011 Aug. 30 | 4, 8, 12 | Custom HCD | Ar | 40 | 2200-3900 | 250,000 | 20 | 60 |
| 39-43 | 2011 Sept. 22 | 2, 5, 9, 12, 16 | Commercial HCD | Ar | 10 | 2200-3900 | 250,000 | 10 | 180 |

| 44-48 | 2011 Sept. 29 | 2, 5, 9, 12, 16 | Commercial HCD | Ar | 10 | 2200-3900 | 250,000 | 10 | 180 |
| 49-53 | 2011 Oct. 24 | 2, 5, 9, 12, 16 | Custom HCD | Ar | 40 | 2200-3900 | 250,000 | 20 | 60 |
| 54-58 | 2011 Nov. 4 | 2, 5, 9, 12, 16 | Custom HCD | Ar | 80 | 2200-3900 | 250,000 | 20 | 60 |
| 59-63 | 2012 Feb. 21 | 2, 5, 9, 12, 16 | Custom HCD | Ar | 40 | 2200-3900 | 250,000 | 40 | 30 |
| 64-68 | 2012 Feb. 23 | 2, 5, 9, 12, 16 | Custom HCD | Ne | 40 | 2200-3900 | 250,000 | 40 | 30 |

[1] At least 3 CCD frames are needed to capture a complete echelle grating order in the UV. Several techniques are used to bridge from one CCD frame to adjacent frames. In some of the above data 5 CCD frames were used to provide redundancy and a check for lamp drift. In other data short exposures were used to check for lamp drift.

[2] Lamp types include commercially available small sealed Hollow Cathode Discharge (HCD) lamps typically used in atomic absorption spectrophotometers and a custom water-cooled HCD lamp.

Table 3. Comparison of experimental branching fractions for 364 lines of Ti II from upper odd-parity levels organized by upper level.

| Wavelength in air[1] (Å) | Upper Level Energy[2] (cm⁻¹) | J | Lower Level Energy[2] (cm⁻¹) | J | Branching Fraction Pickering et al. (2001) | Bizzarri et al. (1993) | This experiment |
|---|---|---|---|---|---|---|---|
| 3383.7584 | 29544.454 | 2.5 | 0.000 | 1.5 | 0.7937 +- 6% | 0.777 +- 1% | 0.78071 +- 0% |
| 3394.5722 | 29544.454 | 2.5 | 94.114 | 2.5 | 0.1533 +- 6% | 0.158 +- 5% | 0.15849 +- 1% |
| 3409.8084 | 29544.454 | 2.5 | 225.704 | 3.5 | 0.0057 +- 5% | 0.00699 +- 5% | 0.00663 +- 2% |
| 3491.0494 | 29544.454 | 2.5 | 907.967 | 1.5 | 0.0366 +- 4% | 0.0448 +- 5% | 0.04121 +- 1% |
| 3500.3331 | 29544.454 | 2.5 | 983.916 | 2.5 | 0.0041 +- 5% | 0.00497 +- 6% | 0.00477 +- 3% |

Notes. –Table 3 is available in its entirety via the link to the machine-readable version above.

[1]Wavelength values computed from energy levels using the standard index of air from Peck & Reeder (1972).

[2]Level energies, parities, and J values are from Saloman (2012).

Table 4. Experimental atomic transition probabilities for 364 lines of Ti II from upper odd-parity levels organized by increasing wavelength in air.

| Wavelength in air[1] (Å) | Upper Level Energy[2] (cm$^{-1}$) | J | Lower Level Energy[2] (cm$^{-1}$) | J | Transition Probability (10$^6$ s$^{-1}$) | log$_{10}$(gf) |
|---|---|---|---|---|---|---|
| 2325.0627 | 47625.048 | 2.5 | 4628.657 | 2.5 | 0.73 +- 0.09 | -2.45 |
| 2348.4066 | 47466.748 | 3.5 | 4897.718 | 3.5 | 0.71 +- 0.10 | -2.33 |
| 2474.1944 | 40798.433 | 3.5 | 393.446 | 4.5 | 0.52 +- 0.06 | -2.42 |
| 2477.2025 | 40581.630 | 2.5 | 225.704 | 3.5 | 0.49 +- 0.06 | -2.57 |
| 2478.6965 | 40425.718 | 1.5 | 94.114 | 2.5 | 0.48 +- 0.06 | -2.75 |

Notes. –Table 4 is available in its entirety via the link to the machine-readable version above.

[1]Wavelength values computed from energy levels using the standard index of air from Peck & Reeder (1972).

[2]Level energies, parities, and J values are from Saloman (2012).

Table 5. Solar Photospheric Titanium Abundances from Individual Ti II Lines

| Wavelength in air (Å) | Lower Energy (eV) | $\log_{10}(gf)$ | $\log_{10}(\varepsilon)$ |
|---|---|---|---|
| 3263.683 | 1.164 | -1.14 | 4.99 |
| 3276.772 | 1.179 | -0.89 | 4.94 |
| 3276.992 | 0.122 | -2.44 | 4.94 |
| 3278.288 | 1.230 | -0.26 | 5.02 |
| 3318.023 | 0.122 | -1.07 | 4.97 |

Notes. –Table 5 is available in its entirety via the link to the machine-readable version above.

Table 6. Titanium Abundances from Individual Ti II Lines in HD 84937

| Wavelength in air (Å) | Lower Energy (eV) | $\log_{10}(gf)$ | $\log_{10}(\varepsilon)$ |
|---|---|---|---|
| 2474.194 | 0.049 | -2.42 | 3.17 |
| 2517.431 | 0.135 | -1.50 | 3.20 |
| 2571.032 | 0.607 | -0.90 | 3.15 |
| 2581.711 | 1.083 | -1.58 | 3.30 |
| 2717.297 | 1.130 | -1.49 | 3.17 |

Notes. –Table 6 is available in its entirety via the link to the machine-readable version above.

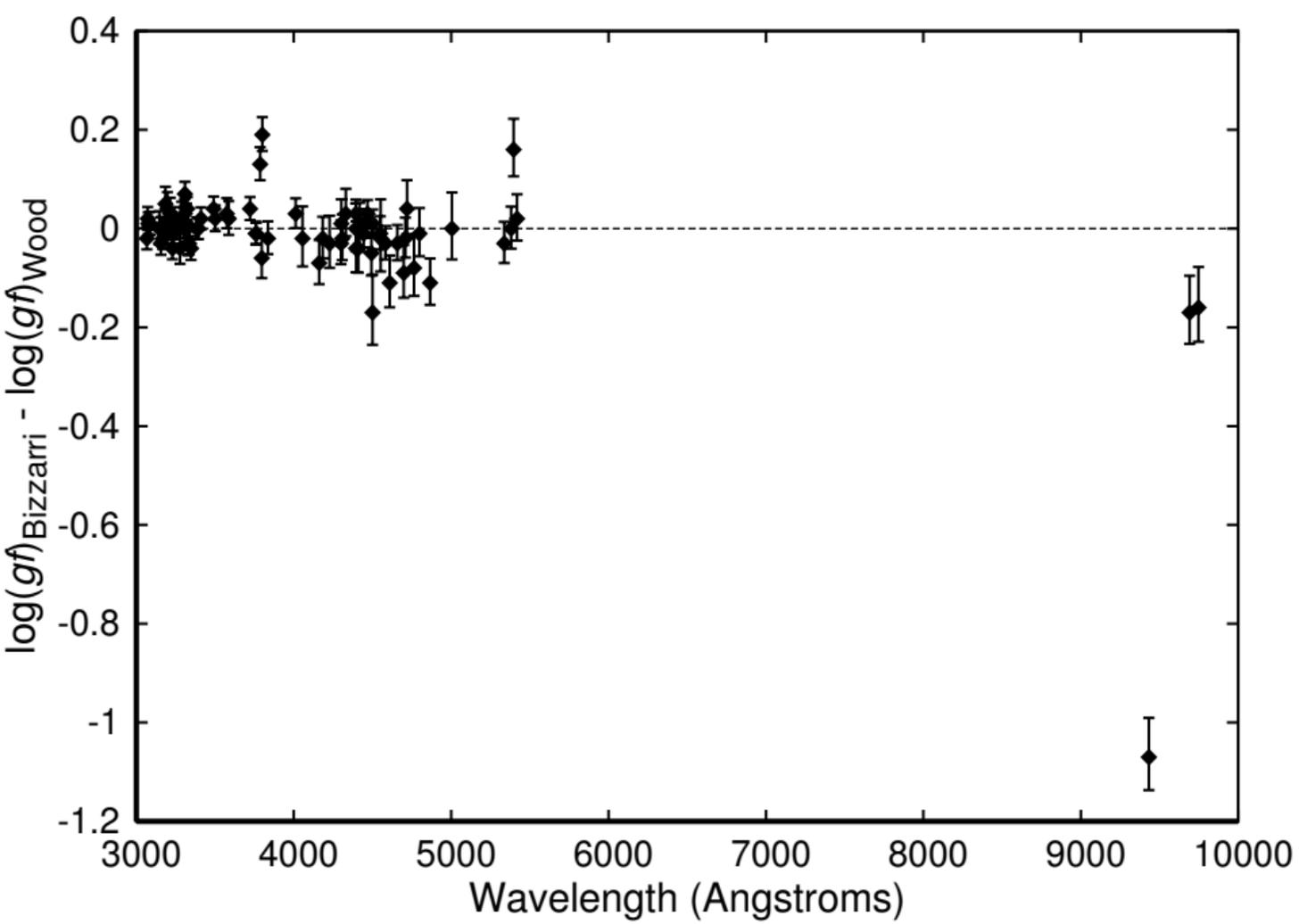

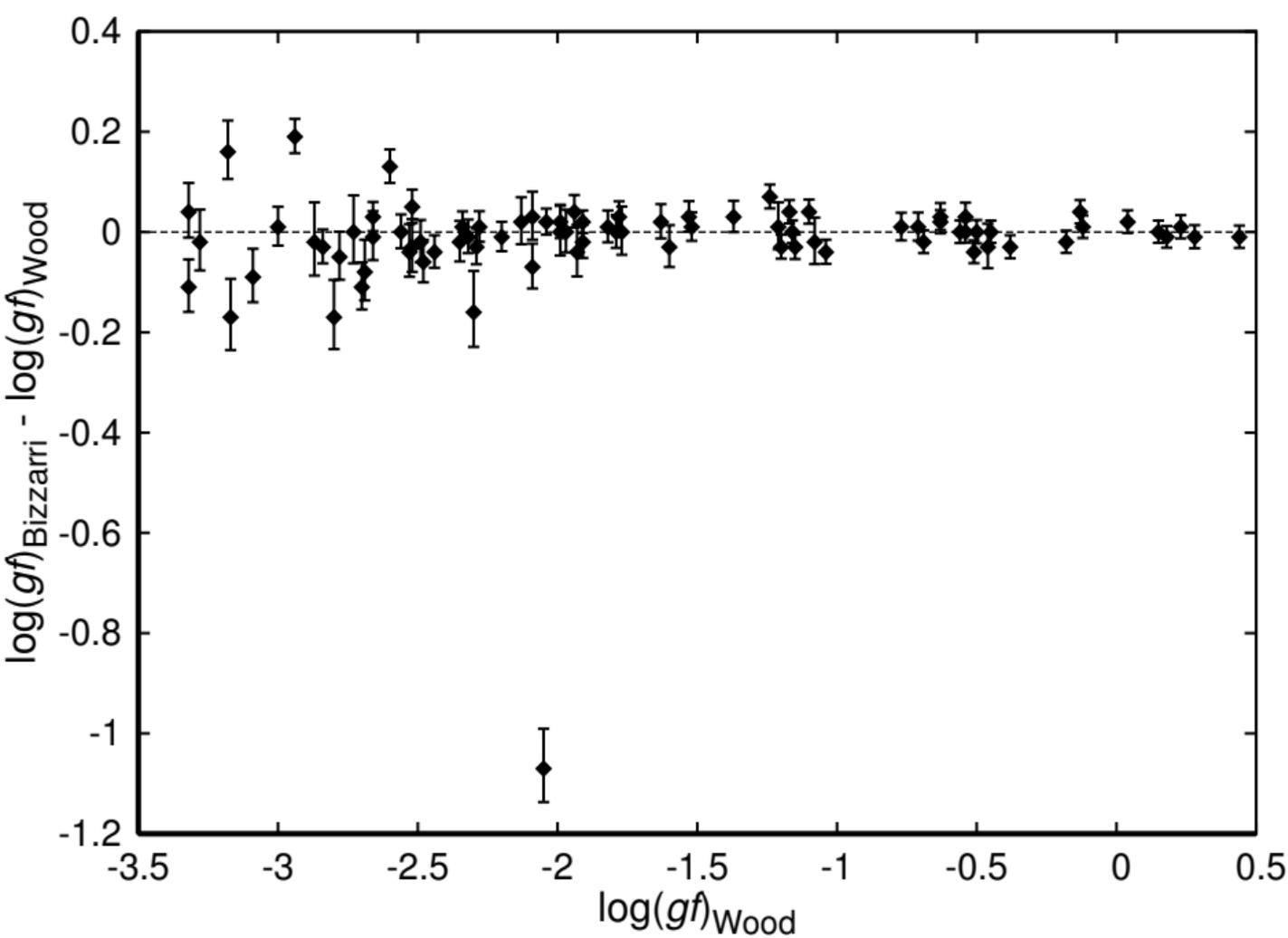

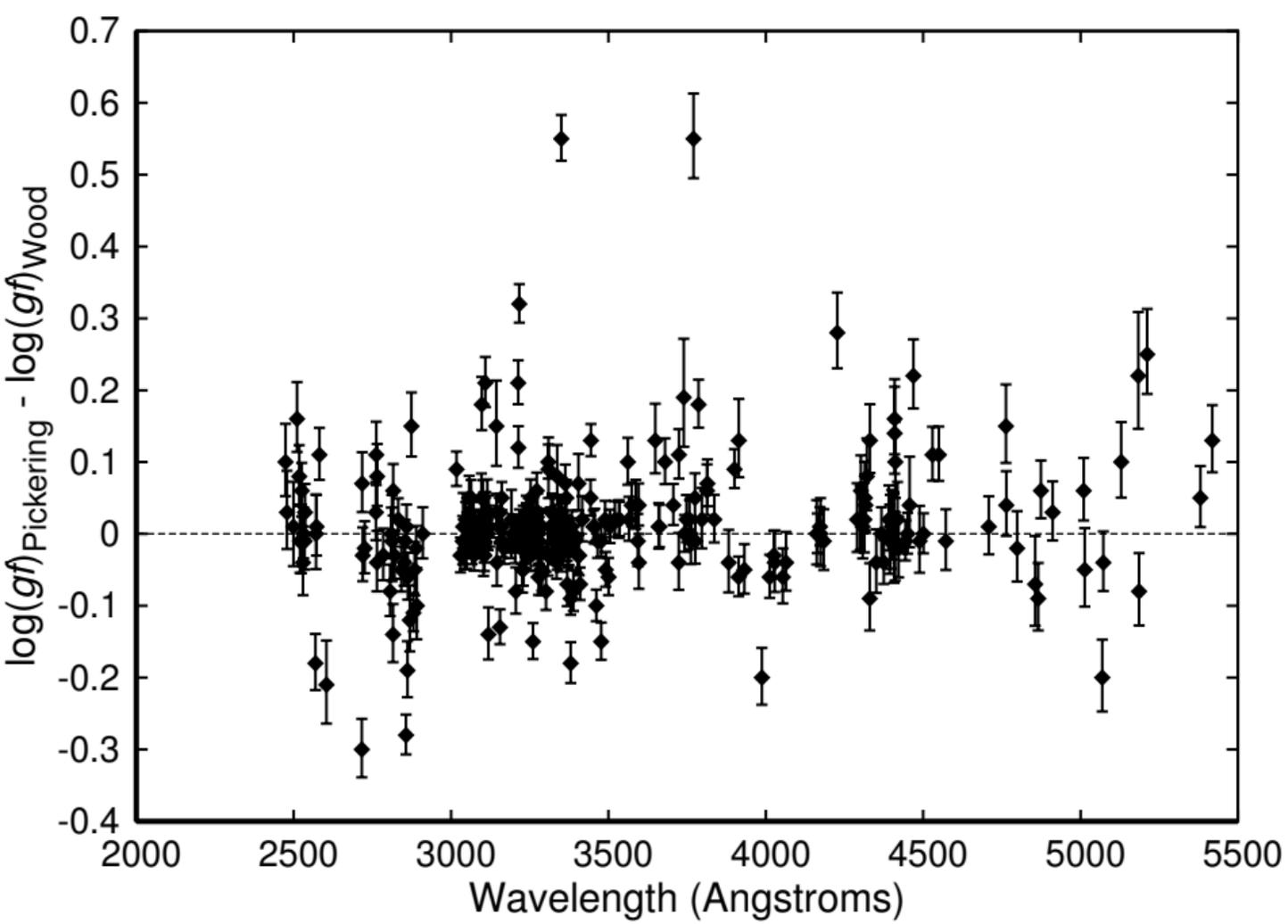

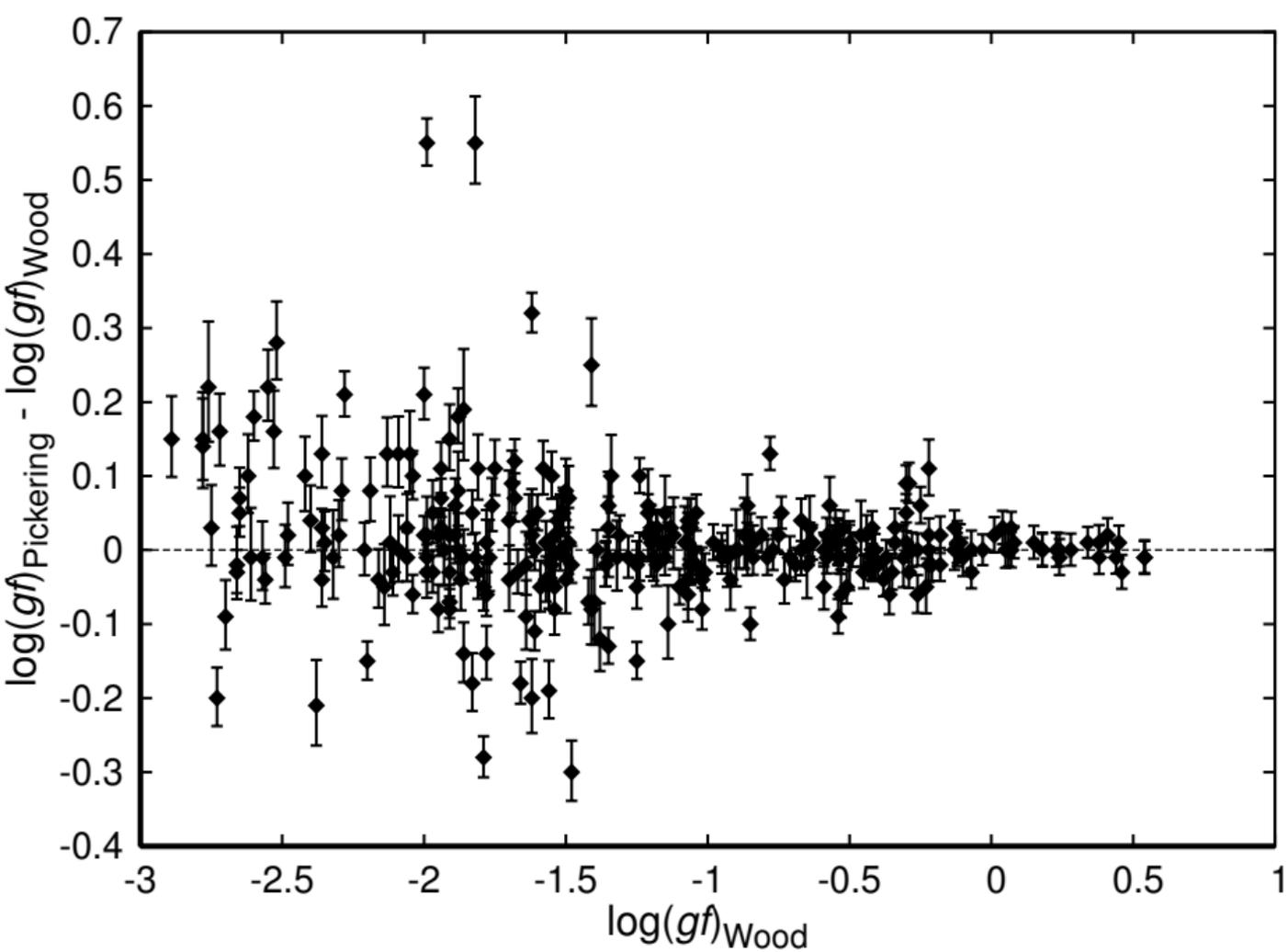

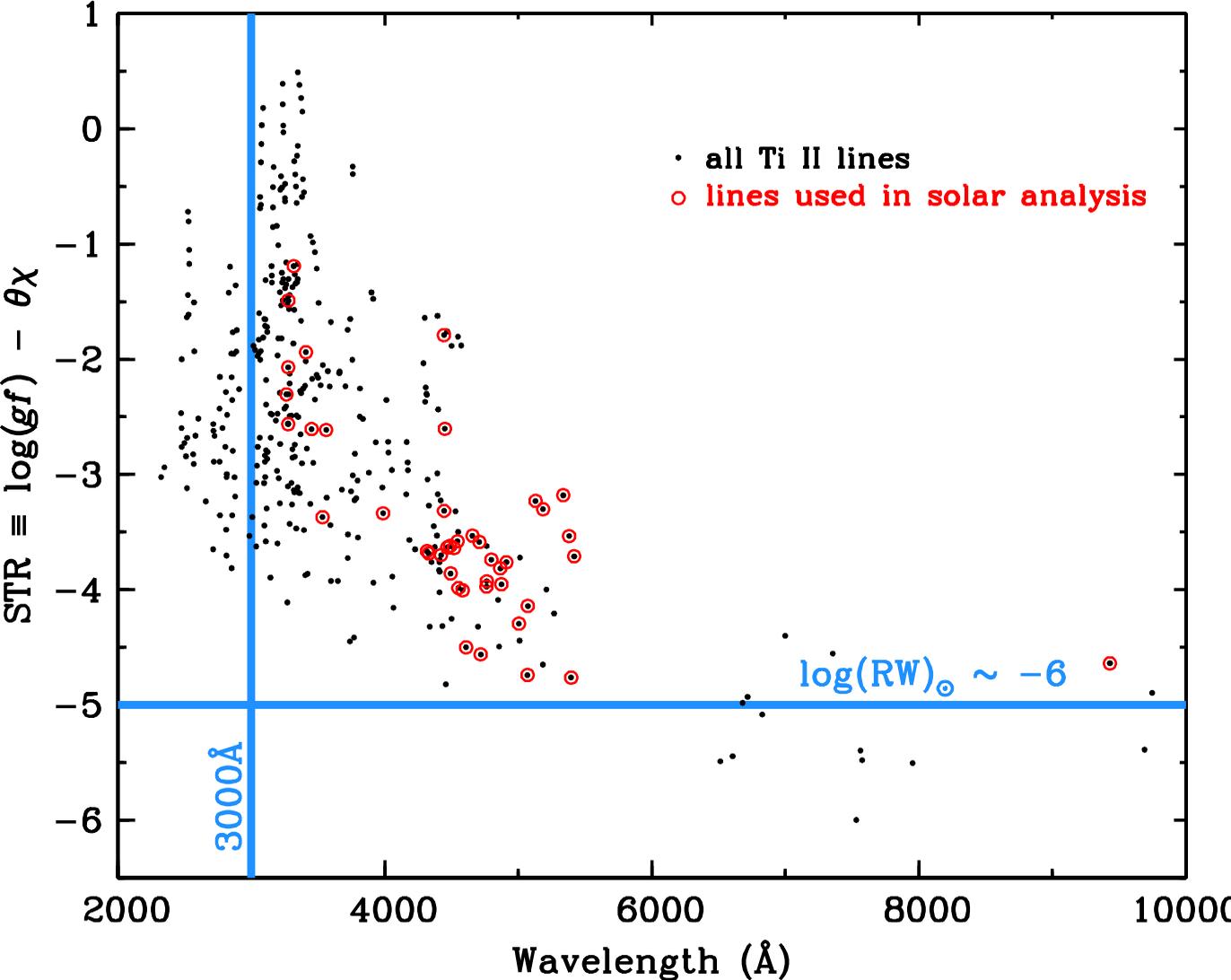

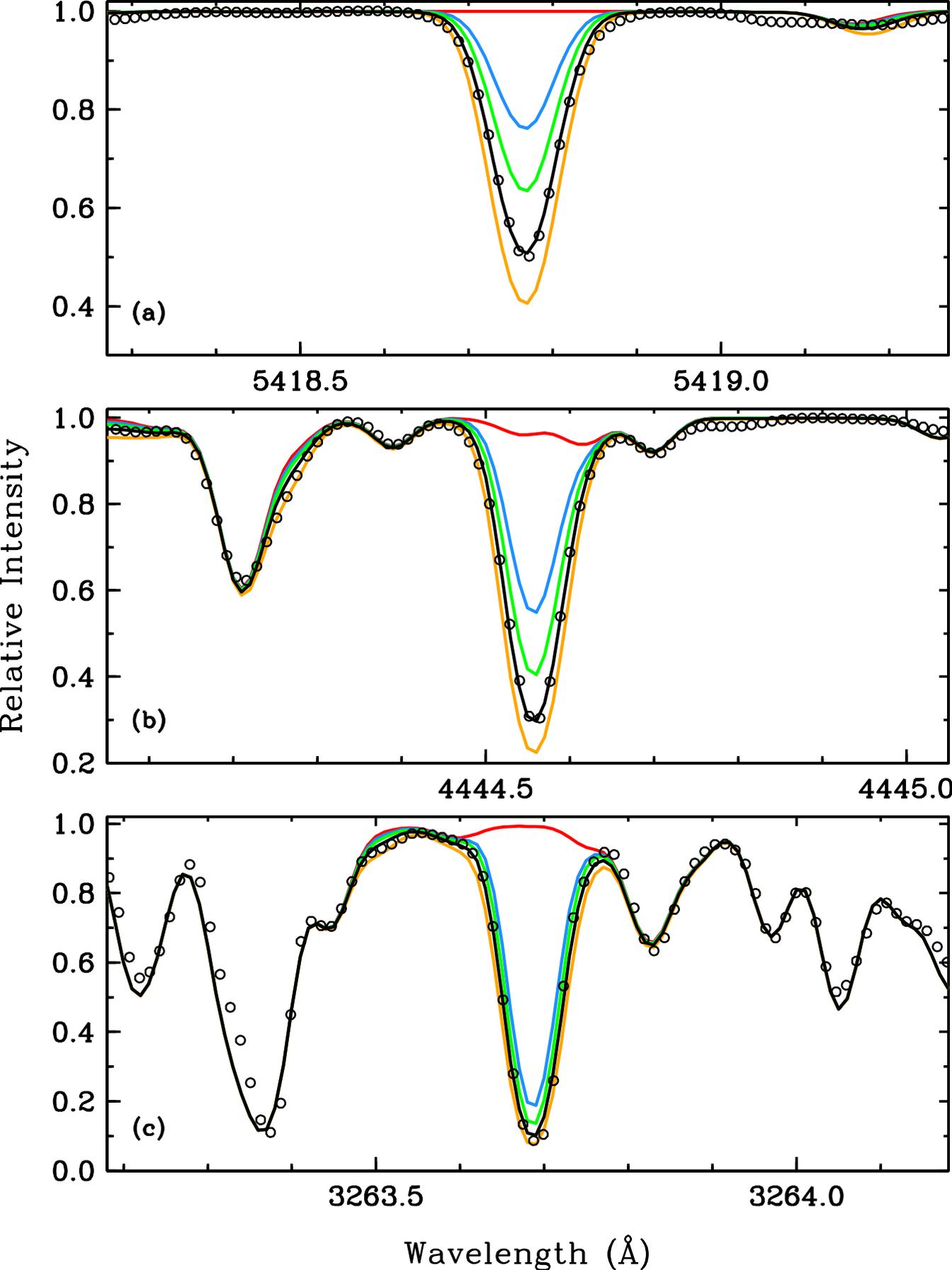

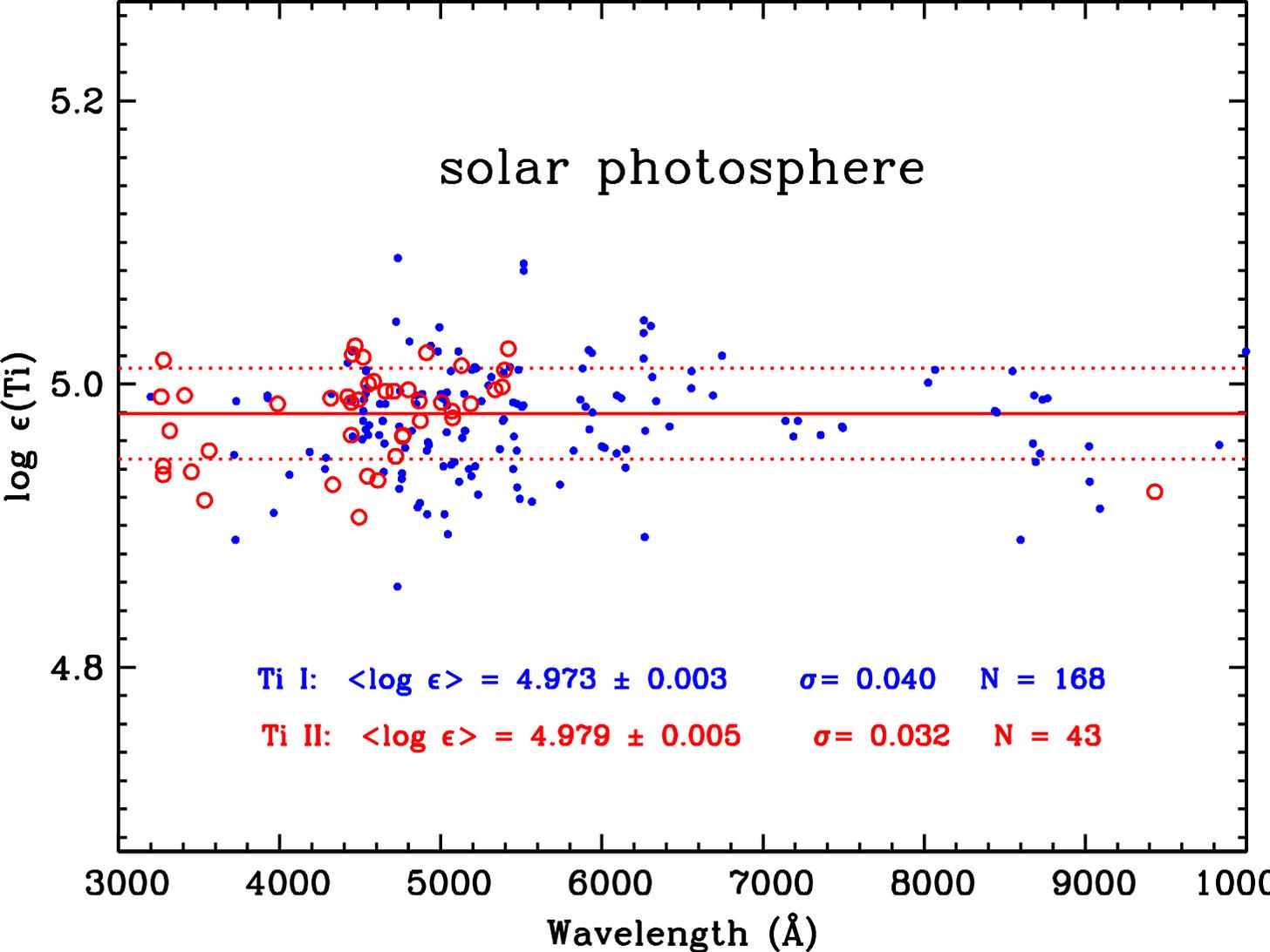

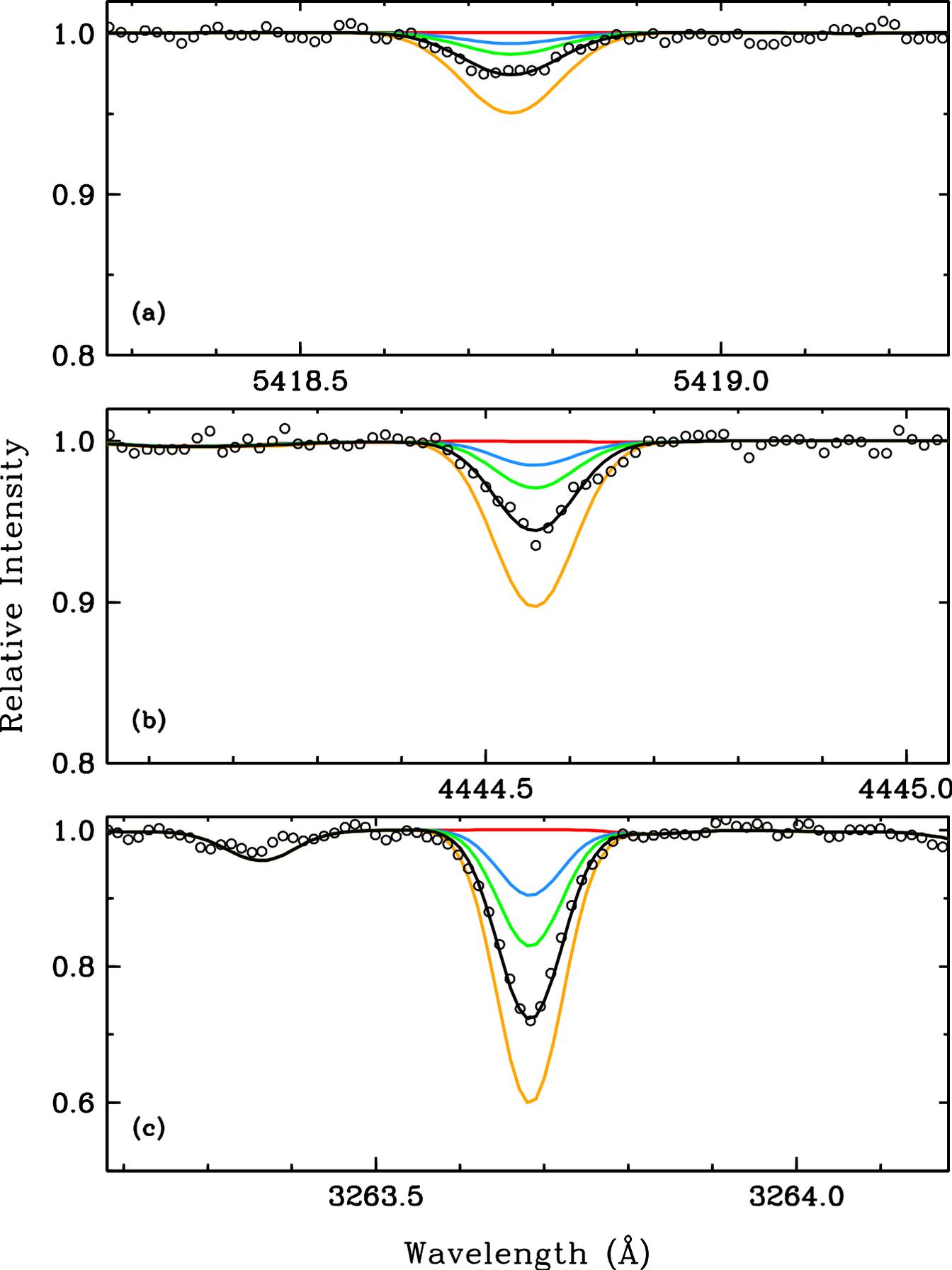

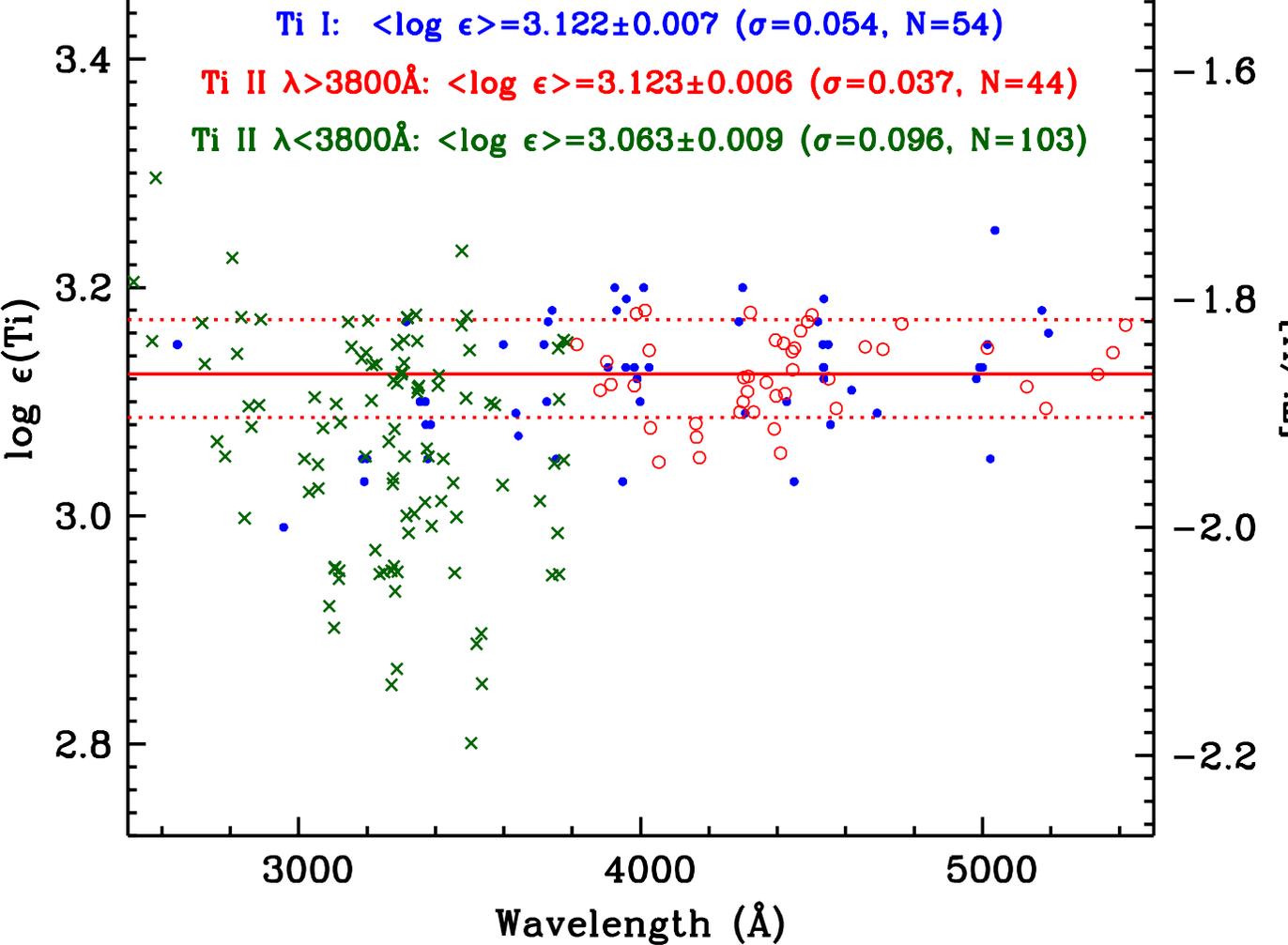

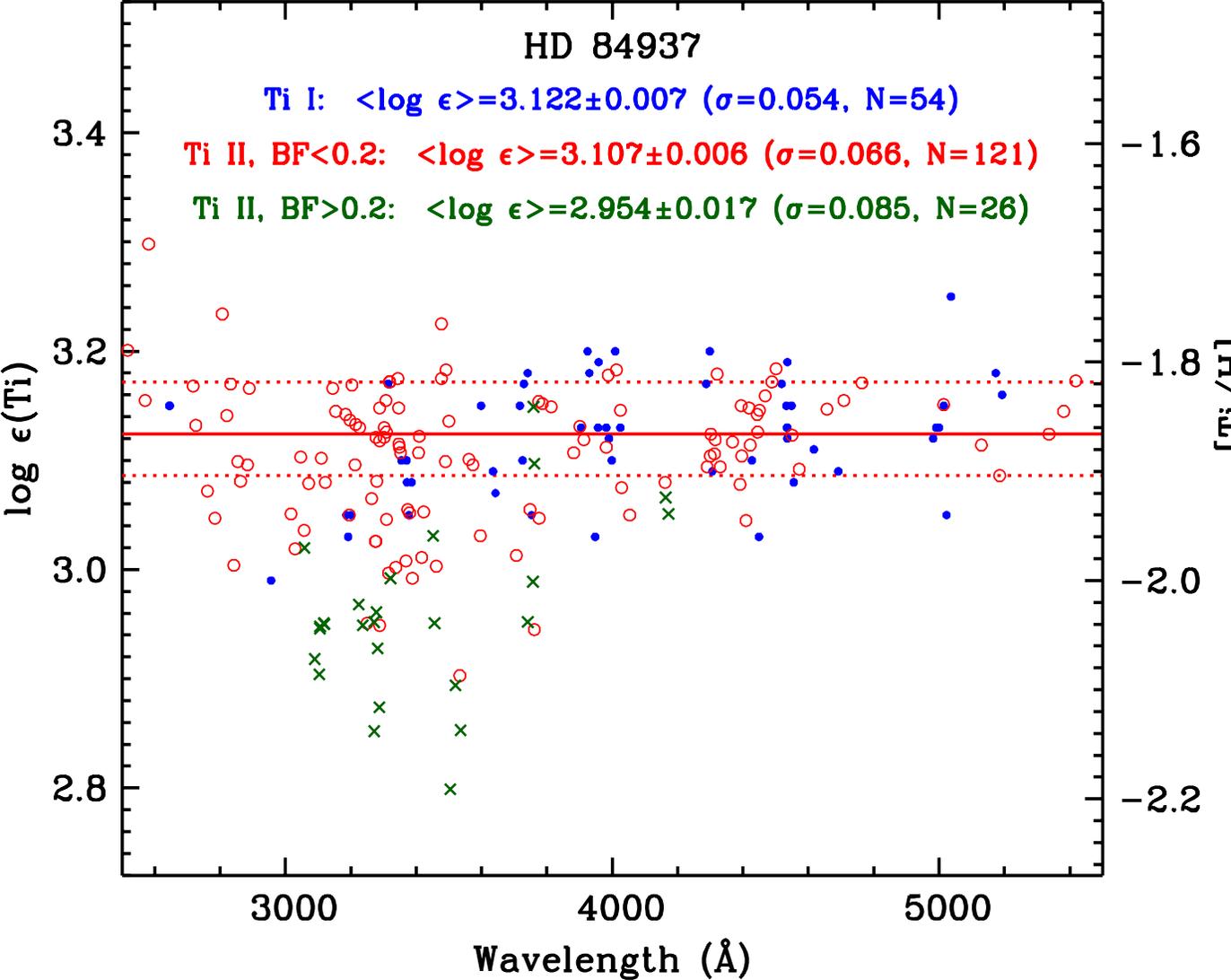

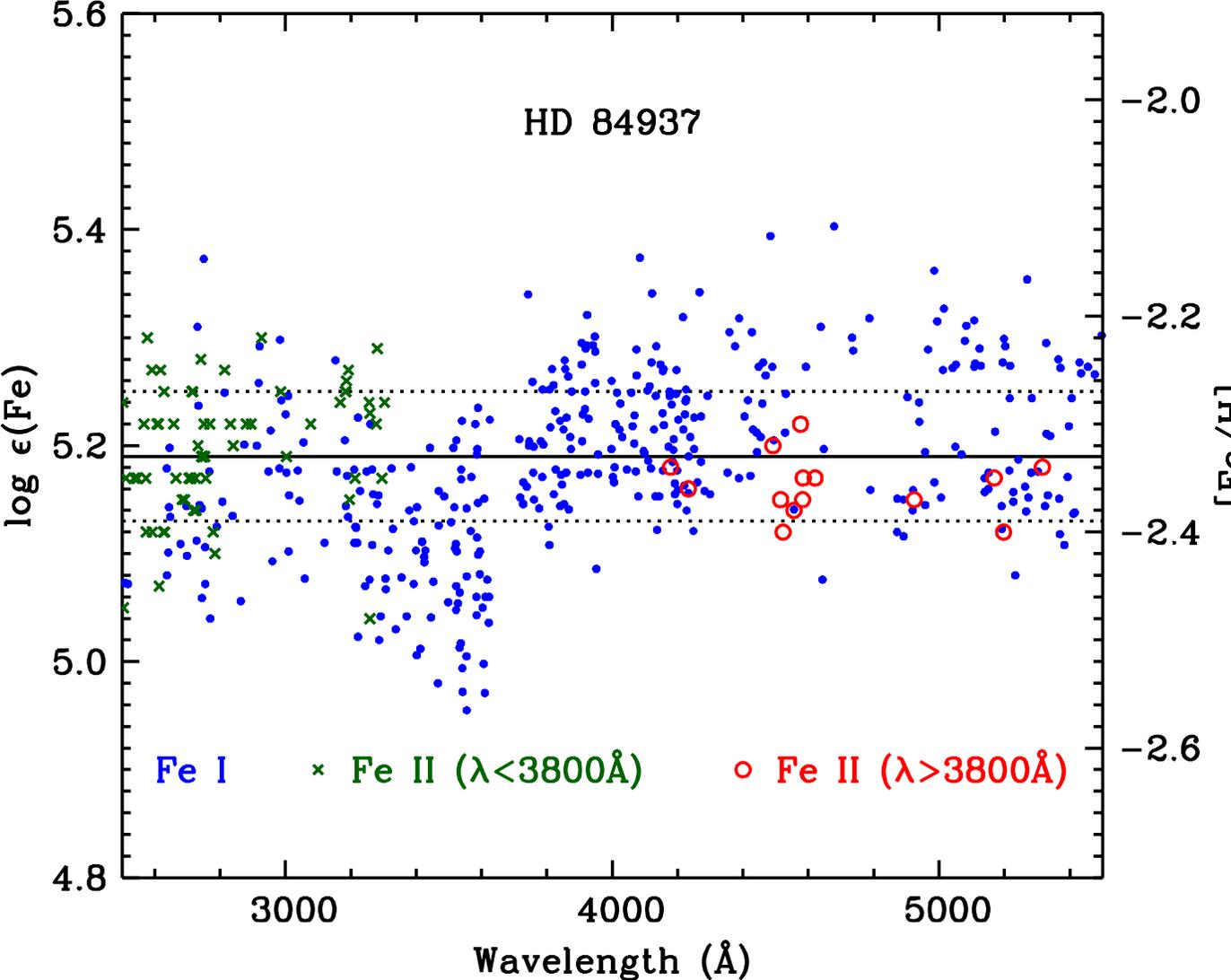

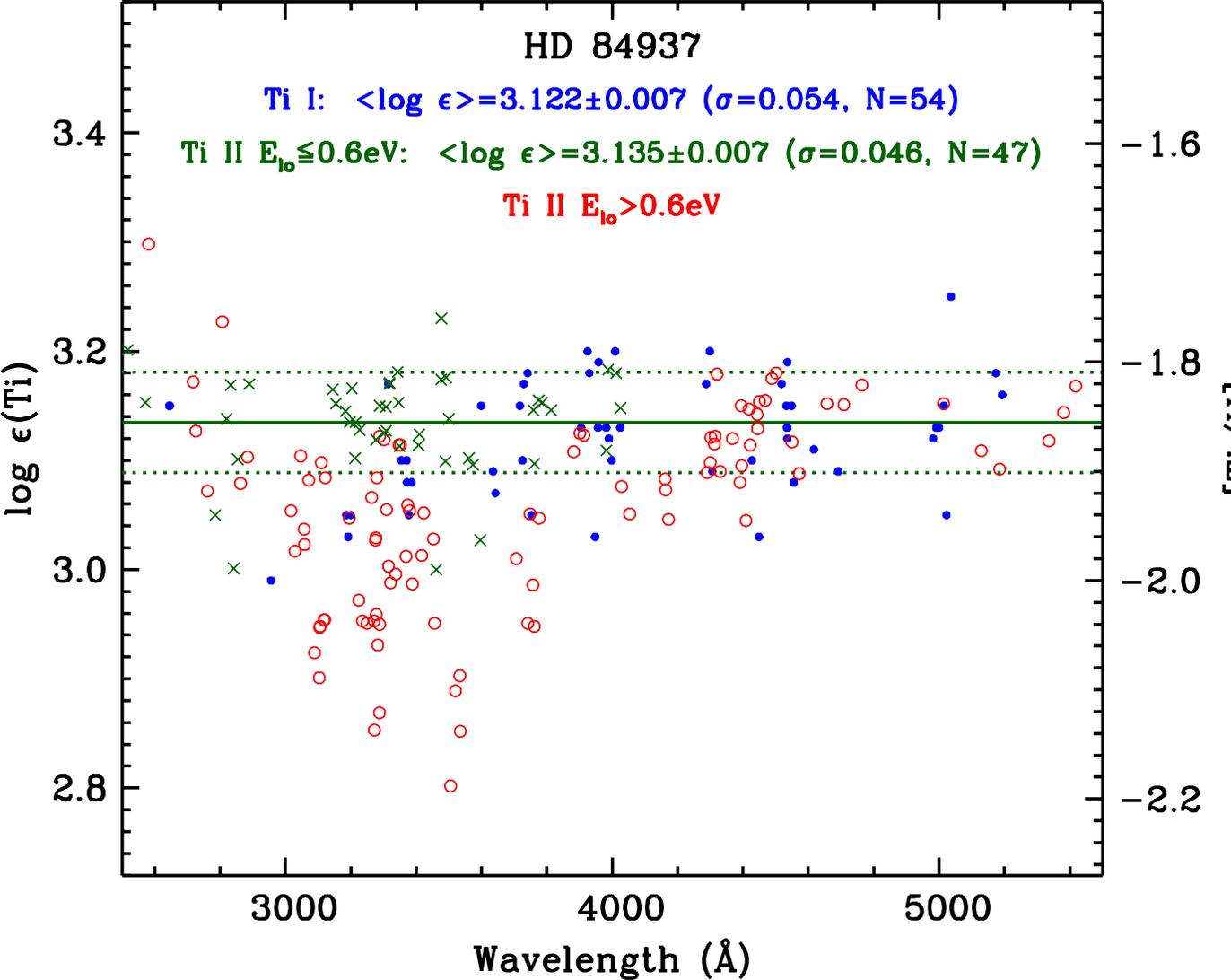